# INFORMATION DYNAMICS SHAPE THE SEXUAL NETWORKS OF INTERNET-MEDIATED PROSTITUTION

Luis EC Rocha[1], Fredrik Liljeros[2] and Petter Holme[1,3]

[1]Department of Physics, Umeå University, 90187 Umeå, Sweden.   [2]Department of Sociology, Stockholm University, 10691 Stockholm, Sweden.   [3]Department of Energy Science, Sungkyunkwan University, Suwon 440-746 Korea

Like many other social phenomena, prostitution is increasingly coordinated over the Internet. The online behavior affects the offline activity; the reverse is also true. We investigated the reported sexual contacts between 6,624 anonymous escorts and 10,106 sex-buyers extracted from an online community from its beginning and six years on. These sexual encounters were also graded and categorized (in terms of the type of sexual activities performed) by the buyers. From the temporal, bipartite network of posts, we found a full feedback loop in which high grades on previous posts affect the future commercial success of the sex-worker, and vice versa. We also found a peculiar growth pattern in which the turnover of community members and sex workers causes a sublinear preferential attachment. There is, moreover, a strong geographic influence on network structure—the network is geographically clustered but still close to connected, the contacts consistent with the inverse-square law observed in trading patterns. We also found that the number of sellers scales sublinearly with city size, so this type of prostitution does not, comparatively speaking, benefit much from an increasing concentration of people.

## INTRODUCTION

Banished from mainstream mass media channels of advertisement, commercial sex is to a large degree organized laterally in social networks among and between sex-buyers and sex-sellers [1]. Over the past decade, the Internet has become an increasingly important vehicle for sharing information about prostitution [2]. With the Internet, sex-sellers have been able to reach other types of customers, those who prefer to establish personal communication before actually buying sex, using the relative anonymity of Internet-based communication [3]. Not only does the flow of information form a network, prostitution also contributes to the web of sexual contacts—the underlying structure by which sexually transmitted infections spread [4]. In this work, we investigated a Web-based community in which sex-buyers rate and comment on their experiences with escorts. The data thus represents a network of sexual encounters and also reflects the information flow organizing a commercial sex scene.

Broadly, this study is related to the recent literature on human dynamics and online social networks (e.g., refs. [5–11]). One theme of these studies is that nontrivial system-wide properties can emerge from interactions in a large population. One example is that although almost all human biological traits follow narrow probability distributions, traits related to social activity (response times in communication, words in texts, wealth, etc.) can be very broadly distributed, so that some individuals have much higher values than average. Such phenomena are usually explained by feedback mechanisms in the interaction between agents via mechanistic models [6–8, 12, 13]. In our data, we can monitor couplings both from offline to online activities and vice versa. We can also study effects of location, space, and urbanity.

We investigated a forum-like Brazilian Web community intended for information exchange between anonymous, heterosexual, male sex buyers. Community members post about encounters with escorts (later we will also refer to them as sex-workers, sex-sellers, or just sellers). From these posts we created temporal networks of the claimed sexual activity. The visible information contains anonymous user nicknames of sex-sellers and -buyers, in which city the activity occurred, and the time of posting (which we took as a rough estimate of the actual time of the sexual encounter).

### The community

The community we studied is a Brazilian, public online forum with free registration, which is financed by advertisements. In this community, male members grade and categorize their sexual encounters with female escorts, both using anonymous nicknames. The forum is oriented to heterosexual males. It is separated into sections corresponding to different cities (or regions) in Brazil and sections about miscellaneous subjects of interest to the members (e.g., tips about hotels, social events, Websites, sexual fetishism). For each city, there is a set of subsections about different commercial sex activities such as escorts, street and brothel prostitution, massage parlors, swingers' clubs, etc. Sometimes, these activities are grouped together in the same section. Every subsection has a number of topics that are either related to a specific escort, club, red-light district, brothel, or another subject. We extract our dataset from the escort section (in Portuguese *acompanhantes*, the word "prostitute" is usually avoided—they are also identified as "free-lancers" or "independent" in the community jargon). There are no regulations about buying and selling sex in Brazil, but pimping, trafficking, running brothels or in other ways facilitating prostitution is illegal [14]. The escorts usually advertise on Websites and sometimes in newspapers (typically as a "masseuse") and offer services in their homes, the customer's home, or hotels. In the taxonomy of Harcourt and Donovan [1], some sex-workers in our dataset can also be classified as "private." Mostly because the escort section is more actively moderated and the escorts grouped into threads, we excluded posts about other types of prostitution (like street prostitution or prostitution in brothels or clubs [1]). The escort's service is identified as higher class and more expensive. Although reading posts is possible without subscribing to the community, registration is needed to post on the forums. Members comment on their encounters and are asked to rate them as bad, neutral, or good. A typical post reports how the escort was contacted (Web-page, telephone, etc.), types of sexual activity (intercourse, oral, etc.), and the sex-seller's attitude and behavior during the encounter, including friendliness and sexual skills. Much attention is paid to the casualness of the encounter, whether the escort acts involved or seems more distant. It is also common to comment on their physical attributes. Community members, on the other hand, are ranked according to their activity on the forum (only measuring the number of posts related to encounters). The moderators claim to pursue a rigorous policy, keeping the forum as reliable as possible. Since it is expected that sex workers neither use real names nor keep the same name over time, the administrator groups the reported aliases of each sex worker together (this is done using photos linked by the members as evidence). This procedure reduces de-duplication of sex-workers though nothing keeps the same individuals from acting as different members. The moderators also remove posts intended to promote or denigrate specific members or escorts; the posts are restricted to encounter reports.

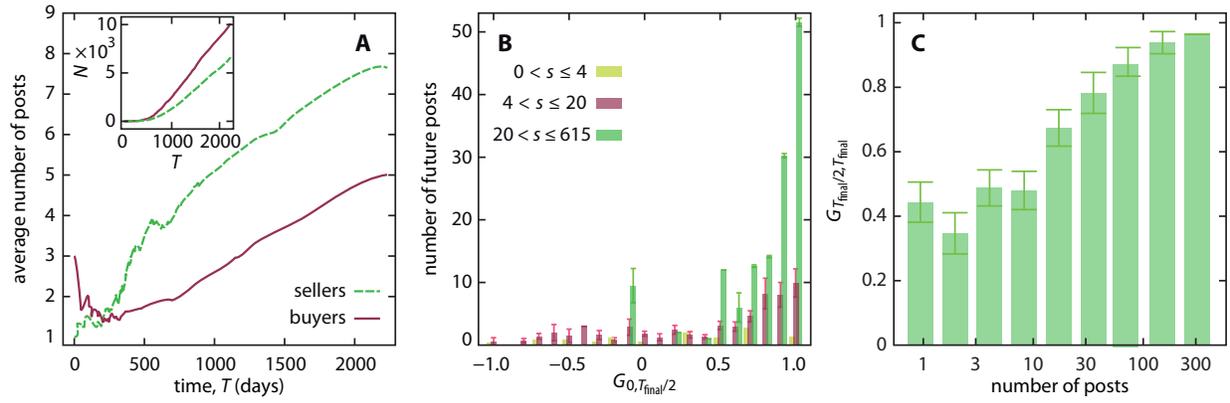

**Fig. 1.** Statistics of the evolution of the community and the feedback from grades to number of posts. (*A*) Time evolution of the average number of posts by sex-buyers and about sex-buyers. The inset shows the growth in the number of sex-sellers and sex-buyers in the data. (*B*) The number of new posts according to the previous average grade at $T_{final}/2 = 1{,}116$ days for three different activity levels, or total number of posts, s. The $R^2$-values of these data are 0.19 ($0 < s \leq 4$), 0.29 ($4 < s \leq 20$), and 0.33 ($20 < s$). (*C*) The average future grade of sellers as a function of their number of contacts at half of the total sampling time (the data is logarithmically binned along the abscissa).

### Network construction and accuracy

From the posts mentioned above, we constructed a network by connecting every community member (sex-buyer) to an escort. To be more precise, an edge between member A and sex-seller B means that A posted a comment in a thread about B. In the analysis, we typically separate the two classes of vertices and talk about properties of either "sellers" or "buyers." We skip posts without a rating of the sex-worker and consider multiple entries separately in case the same two individuals had sex more than once. Furthermore, we save information about the time of the post and the rating of the sex-worker. The data extend from the beginning of the community; they cover the period between September 2002 and October 2008.

The data is self-posted and self-anonymized by the sex-buyers, and publically visible. This, along with the covert nature of prostitution, makes fake or erroneous posts quite possible. In this paper we have not tried to compensate for such effects, but used the raw data and assume the biases are not large enough to invalidate our conclusions. One reason to believe that this is feasible is that the community watches its members putting a social pressure to keep the data correct. This certainly applies to the identity of sex-sellers, as noted above, but also to sex-buyers reviewing faux encounters who run the risk of being discovered by other sex-buyers buying sex from the same escort around the same time. Another reason for a sex-buyer to keep his identity is that the activity level associated to the nickname is a source of prestige. In sum, to fulfill its function as an information source, the community is cleaning itself from errors. Furthermore, none of the statistics we use is more than linearly sensitive to errors. The quantitative conclusions would also tolerate systematic levels of error—if the real-life sex-buyers have, say, 1.5 nicknames and report 30% more encounters on average, but that these levels are fairly stable across the sampling time and location, then the qualitative conclusions would not be affected.

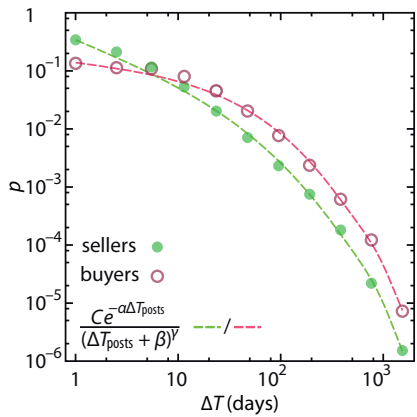

**Fig 2.** Distribution of the time elapsed between two posts $T_{posts}$ for buyers and sellers. Many posts were written during the same day, respectively, $p(T_{posts} = 0) = 0.495$ and $p(T_{posts} = 0) = 0.246$. The distributions are well fitted by $p(T_{posts}) = C \exp(-\alpha T_{posts}) = (T_{posts} + \beta)^\gamma$, with: $C = 2.9 \pm 0.5$ days$^\gamma$, $\alpha = 0.0023 \pm 0.0001$ days$^{-1}$, $\beta = 3.1 \pm 0.4$ days and $\gamma = 1.49 \pm 0.04$ (for sellers); and $C = 12 \pm 8$ days$^\gamma$, $\alpha = 0.0021 \pm 0.0002$ days$^{-1}$, $\beta = 18 \pm 4$ days and $\gamma = 1.5 \pm 0.1$ (for buyers).

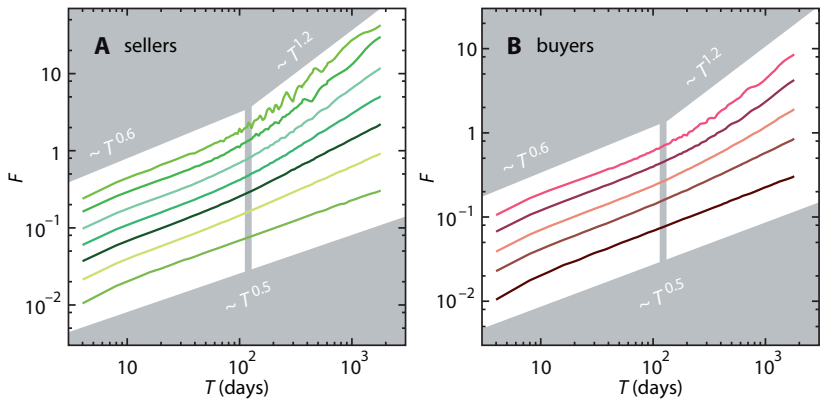

**Fig. 3.** Detrended fluctuation analysis. (*A*) shows the DFA fluctuation function as a function of the timescale Δ*T* for sellers and (*B*) shows the corresponding plot for buyers. The curves in (*A*) and (*B*) corresponds to different activity levels. From bottom to top the curves correspond to less than 3, 3–7, 8–20, 21–54, 55–148, 149–403, and more than 403 posts (about sellers or from buyers) respectively. Black lines are inserted for reference. $t^{1/2}$ corresponds to uncorrelated interaction.

## RESULTS

In this section we discuss the time evolution of the forum and its emergent spatiotemporal patterns and network structure.

### Growth of the community

In general, the growth of a forum-based Web community consists of three basic processes: new community members become active; community members start new topics (threads); or community members post in old threads. Members may also quit a community, but in actual datasets (in our study as well) that is often indistinguishable from an extended hiatus. Specific to our forum, one thread corresponds to one escort, so any post is either a community member writing about a new escort not previously in the community, or in an old thread about an already discussed sex-worker. In the first case, a new vertex (the sex-seller) is added to the network, as is an edge connecting the sex-worker and the member. In the second case, if the community member is posting about that sex-worker for the first time, a new edge will be added. We plotted the growth of the two types of vertices, from the beginning of the Website (*T* = 0) up to the end of our sampling ($T_{final}$ = 2,232 days) in Fig. 1*A* (inset). After an initial period of about 500 days, the membership grew at a fairly constant rate of about six new members per day. The number of sex-workers in the data show a similar time evolution but with a slightly lower growth rate in the latter phase of about five individuals per day. If the sampling time had been longer, the average number of posts per individual would probably have converged. As seen in Fig. 1*A*, the average (vertex) degree (number of network neighbors) was growing for both buyers and sellers. This means that our sampling time was shorter than the typical time that a sex-buyer or seller had an active presence in the forum. The curve for the sellers shows an incipient tendency toward saturation. This incipient saturation suggests that the timescale of a sex-seller's "career" is not much longer than the sampling period, and also that sex-buyers stay in the commercial-sex arena longer than do sex-sellers. For *T* < 250, the average number of posts is larger for the buyers than sellers, suggesting an early core community that later gets diluted by less active members.

### Grades and activity of sellers

From Fig. 1*A* it is hard to see traces of the feedback processes between the encounters (the sexual network) and the information processes in the community. A strong candidate channel for feedback is the rating in the posts. The grade functions as a consumer advice to other sex-buyers and adds a more objective flavor to the textual comments. The grade is given by checking a box described only by a word. To get an accumulated score, we assigned numerical scores as follows: −1 for "bad" encounter, 0 for "neutral" and +1 for "good." Then we considered the average value of these scores $G_T$ from the beginning of the data set to time *T*. We tested our assumption that this average score is correlated with the sex-worker's ability to acquire new customers in Fig. 1*B*. For a given time $T = T_{final}/2 = 1,116$ days, the number of community members posting about a specific sex-worker increases by $G_T$ (Fig. 1*B*). We separated the analysis, considering three different intervals of the total number *s* of posts about the sex-workers. Highly rated sellers attracted more new customers; whether the seller had a mid or low grade does not seem to have mattered as much. As it turns out, the reverse is also true (Fig. 1*C*)—an escort's future score increases with the number of past contacts, more than can be explained by a time-independent correlation between grade and degree. There is thus a full feedback loop, from the online grades to the offline commercial activity and back to the online behavior.

### Interevent time statistics

The above results describe the behavioral pattern of community members and sex-workers over the history of the community. Now we turn to a more detailed picture of the community's dynamics. In Fig. 2 we have plotted the probability distribution of the time between two subsequent posts $\Delta T_{posts}$ (cf. Refs. [6, 7]). The distribution is narrower than a power-law, but wider than an exponential or Poisson distribution, and consistent with a power-law with an exponential cut-off $P(\Delta T_{posts}) = C \exp(-\alpha \Delta T_{posts}) / (\Delta T_{posts} + \beta)^\gamma$ (see the *Supplementary Information, SI*). That the distribution is broader than Poisson suggests a more complex dynamic than just sex-buyers posting independent of each other; that it is not a power-law means that the problem cannot be mapped onto other studies of human response dynamics [6, 7]. One explanation for this broad distribution is the feedback loop of community information affecting the sexual activity that then results in new posts (information) on the forum. If this is true, it also means that the sex-seller's workload is affected by forum activity. Another scenario would be that the dynamic is driven by a complex social-situation of the sex-sellers, where their work as escorts is intermittent. The curve for sex-sellers is steeper than the curve for sex-buyers. This can be explained as a collective dynamic phenomenon—one post about a sex-worker triggers another post soon thereafter. Or it could reflect that sex-workers are, at times, more active than buyers (as they have economic incentives to be active, while the buyers benefit from abstinence as they are, in practice, paying for their degree). The broad $\Delta T_{posts}$-distribution can have implications for disease spreading as it suggests wide fluctuations in the number of concurrent sex partners [15] affecting the time ordering of contacts [16–19] and thus the possible pathways of pathogen transmission. We will mention some time-ordering statistics below but not focus on such dynamic aspects.

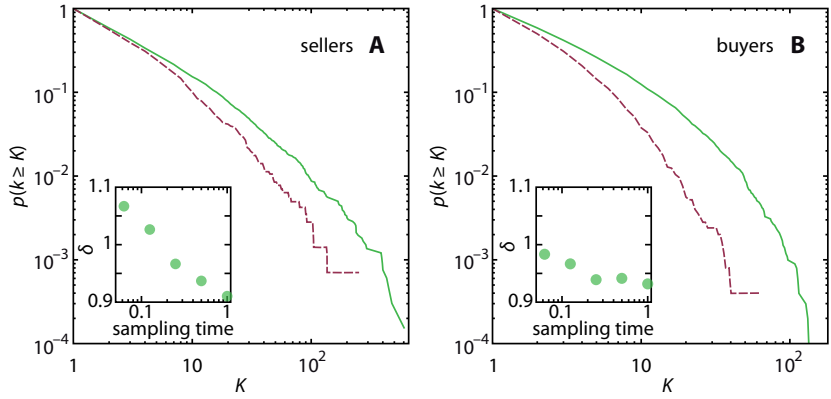

**Fig. 4.** Degree distributions. (*A*) sex-sellers and (*B*) sex-buyers cumulative degree distributions for the full sampling time (solid line) and a yearlong window (starting one year after the full dataset; dashed line) for sex-sellers and -buyers respectively. The insets show the exponent of preferential attachment (cf. Eq. 2).

To get another view of long-range correlations in the data we follow the approach in Ref. [8] and map the dynamics to a random walk. For a buyer or seller $i$ at time $T$, define $Y_i(T)$ as the deviation from the expected accumulated number of messages up to that point, given the average rate of posts over $i$'s presence in the data. For example, if a buyer $i$ posts on average one message per day, but posted ten messages during his first five days in the data set, then $Y_i$(5 days) = 10 – 5 = 5. Next, define $F_i(\Delta T)$ as the mean-root-square of $Y_i(T - \Delta T) - Y_i(T)$, thus quantifying the fluctuations of $Y_i(T)$ over a time scale $\Delta T$. Here we also remove trends in the data by a "detrended fluctuation analysis" described in detail in Ref. [8]. If the sex-buying behavior is random, $F_i(\Delta T)$ will be proportional to $(\Delta T)^{1/2}$. In Fig. 3, we plot the average values of $F_i(\Delta T)$ for buyers (*A*) and sellers (*B*), and different activity levels. Both the curves for sellers and buyers with a large presence in the data show anomalous long-range correlations over about 110 days. These correlations cannot be explained by random sex-buying behavior— the active buyers tend to have long-term customer relationships with escorts. For times less than 110 days, these connections are drowned in a more random sex-buying pattern. Other datasets of online communication (not directly related to offline contacts) do not show the different regimes, but show anomalous correlations for the most active individuals over a larger time span [8] (the only other timescale we find is a one-week pattern, see the *SI*).

## Preferential attachment

Several studies have reported a large variation in claimed number of sexual partners [19–22]. One explanatory mechanism for this large variation in turnover rate is preferential attachment, that is, that having had many previous encounters increases the probability of having more in the future [13, 23]. Preferential attachment is usually an indirect mechanism; in our case it can occur due to the feedback mechanisms discussed in the previous section. To evaluate to what extent the partner turnover rate for sex-buyers and sex-sellers is governed by preferential attachment, we have estimated the extent to which the rate of partner turnover, defined as the probability that the next contact will attach to a vertex, increases linearly with the vertex degree of the sex buyers and sex sellers by fitting $\delta$ in

$$\mathrm{Prob}[k_i(T+1) = k_i(T) + 1] = \frac{k_i(T)^\delta}{\sum_j k_j(T)^\delta} \quad (1)$$

with maximum likelihood estimates for different time intervals. Sex-buyers exhibit sublinear preferential attachment for both short and long intervals. We observed close to linear preferential attachment (even slightly superlinear) for sex-sellers for short time intervals (Figs. 3*A* and *B*), while longer time intervals were associated with sublinear preferential attachment (i.e. $d < 1$). This means that feedback processes are stronger for shorter than for longer timescales. Changes in the life situation of the sex-workers are a possible explanation for this loss of long-term feedback. From Ref. [13] we know that linear preferential attachment yields power-law degree distributions. The observed nonlinear preferential attachment with a sampling-time-dependent exponent gives rise to degree distributions better described as stretched exponentials than pure power-laws (Figs. 3*A* and *B*). If we truncate the sampling time, the degree distribution becomes closer to a power-law.

## Tendencies in types of sexual activities

From the data we can also follow the

tendencies in sexual practices that develop over time. As mentioned, the posts also detail the sexual activity (three categories—anal sex, use of condom during oral sex, and kiss on mouth). For a given agent, let $\tau$ be the order of a post in the dataset ($\tau = 1$ for the first post about a seller or a buyer, $\tau = 2$ for the second, and so on), the *post number*. In Fig. 5, we plot the fraction of a reported type of sexual activity over all of the sellers or buyers, averaged over posts at occasion $\tau$, as a function of $\tau$. We also divide the data into different activity levels (corresponding to the most, middle and least active of the respective sets of agents) to see if trends can be connected to any of these groups. For the sellers, the frequency of all of these sexual services increases with time spent in the data set and with the activity. Most conspicuously, the oral sex without condom increases from 60–75% for sellers as they enter the dataset, to over 95% for the largest $\tau$. There are similar increasing tendencies among buyers too, but except for oral sex without condom, considerably weaker. The most active sellers are also the ones selling the extra services during the largest percentage of encounters. Assuming that this increase in additional services is a general phenomenon (also implying more risky behavior with respect to disease spreading), it would be interesting to investigate the causes in future studies. Is the mutual increase for both sellers and buyers a result of a feedback loop between sellers and buyers? Or does it reflect a natural development of the sellers and buyers, independent of the on- and off-line interaction?

**Large-scale network structure and implications for disease spreading**

The growth mechanisms investigated in the previous sections generate a complex structure of connections between sex-buyers and sex-sellers. This sexual network can be an underlying structure over which information flows (of the type of sex performed or other arrangements during the encounter, for example), but also (and perhaps more importantly) disease. The structure of the network is an important factor determining the large-scale behavior of these dynamic systems [4, 12, 24]. Much focus has been on the probability distribu-tion of degree, especially the observation that vertices with highest degree that are influential in the epidemics (of disease or information) since they have both a higher chance of becoming infected and of infecting others [4]. Other network structures—not just how many neighbors the vertices have, but how the edges are wired on a large scale—could also affect disease spread and give information about the system's social organization. One such quantity is assortativity, which quantifies the correlation between the vertices connected by an edge. Positive assortativity implies that the large-degree vertices are typically connected to each other, while low-degree vertices are linked to other low degree vertices. This means that assortativity measures the tendency of vertices of a similar degree to connect with each other, while a negative value signals that edges between vertices with different degrees of magnitude predominate [12, 24]. An interesting implication for disease spread is that in assortative (artificial) networks outbreaks happen more easily but do not become very large, whereas in disassortative networks epidemics are rarer, but if they occur they spread to a larger part of the network [25]. We observed a small disassortativity (see Table 1). In our context, it means that more active members had a tendency to buy sex from less active sex-workers and, symmetrically, that fewer active buyers could be connected to popular sex-workers. Other studies of Web communities [5] have observed similar characteristics, that is, small but significant disassortativity.

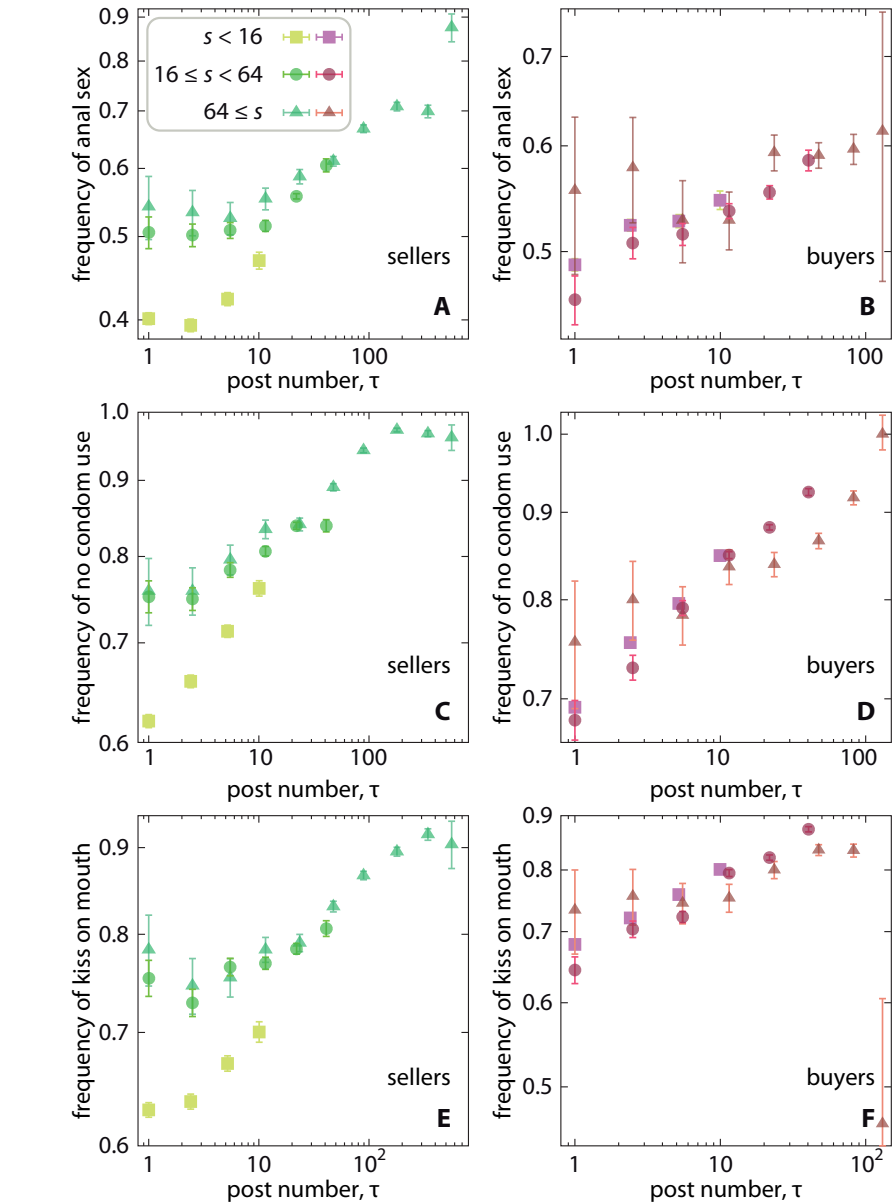

**Fig. 5.** The frequencies of different additional sexual services performed as a function of the time of presence in the data (measured by the post number $\tau$, i.e., the order in the sequence of posts about a seller or by a buyer). (A) and (B) show the frequencies of reported anal sex for the sellers and buyers respectively. (C) and (D) show the statistics for no condom use during oral sex, and (E) and (F) for kisses on the mouth. The error bars show standard errors.

| metrics | buyers | sellers |
|---|---|---|
| no. of vertices | 10,106 | 6,624 |
| no. of vertices in g.c. | 9,652 | 6,158 |
| no. of edges | 40,895 | |
| no. of encounters | 50,185 | |
| | original | randomized |
| diameter of g.c. | 17 | 13.2 ± 0.1 (s.e.) |
| avg. distance | 5.78 | 4.921 ± 0.002 (s.e.) |
| no. of 4-cycles | 231,439 | 64,360 ± 302 (s.e.) |
| assortativity | −0.110 | −0.0896 ± 0.0005 (s.e.) |

**Table 1.** Network metrics associated to the sexual network: total number of vertices of the different types (g.c. means giant component, the largest connected subgraph); number of vertices in the largest connected component; number of edges; number of encounters; diameter (the largest distance between any pair of vertices in the graph; distance is the number of edges of the shortest path between a pair of vertices) of the largest connected component; average distance; number of cycles of length 4, and assortativity (the tendency to connect to vertices with similar degree, see Ref. [23]); s.e. is the standard error.

Another network structure that both reflects the organization of the community and can be informative with respect to disease spread is the density of short cycles (sequences of adjacent edges that end where they start). Since the network is bipartite, we measured the number of four-cycles (Table 1) and observed an approximate fourfold increase compared to a randomized null-model (random networks with the same degrees as in the original network and only edges between buyers and sellers, no other constraints [24]). This is an even larger overrepresentation than observed in online dating networks [5], and in stark contrast to the under-representation of four-cycles found in off-line dating networks [26]. A high density of triangles is a factor lowering the speed of epidemics and is one of the proposed explanations for the slow (polynomial) growth observed in some disease outbreaks [27].

In acquaintance networks, the density of triangles is high and often attributed to one person's introducing two friends to each other, thereby creating a triangle [12]. We believe this over-representation of four-cycles relates to a geographic effect—the cities create dense network clusters where the probability of four-cycles is much higher than if contacts were made, irrespective of distance (only 3% of the four-cycles involve a buyer away from his hometown).

Another factor for disease spreading is time-ordering effects (disease, or information, cannot spread from A to C via B if all contacts from B to C happens before all contacts between A to B) [16–19]. Even if we consider time ordering, the present data is quite connected. The average number of individuals reachable, following the order of the contacts, from an individual present in largest connected component at the first fourth of the sampling time is 11,272 (which is 71.2% of the largest connected component or 67.4% of the entire system). We can conclude that, a majority of the individuals are connected in such a way that information, or disease, can spread between them. It is of course not a closed system, so this percentage is an underestimate. Getting better estimates than this is a modeling challenge of the future.

### Effects of urbanity and geography

An increasing fraction of the Earth's population lives in cities. One reason for this increase is the nonlinear benefits of close proximity. Bettencourt *et al.* [28] have investigated how different socioeconomic indicators scale with city size and found that quantities related to "social currencies such as information, innovation or wealth, associated with the intrinsically social nature of cities" typically scale superlinearly with city size, while quantities scaling linearly are related to basic needs such as water and household energy consumption, and quantities scaling sublinearly relate to transportation and distribution infrastructures like the surface areas of roads. In Fig. 6*A* and *B*, we show the scaling of the number of sellers and buyers as a function of the population of the city they are most active in. The small number of cities in the data put rather large error bars on the results, but we can say with fairly high confidence that the number of sellers scale sublinearly with the city size, while the buyers are closer to a linear function of the population of the cities they live in. This means that engaging in Internet-mediated prostitution does not benefit as much from the "intrinsically social nature of cities" as do other optional careers. One reason for this is probably that the Internet removes the need for a first face-to-face contact, and with it the increasing return on an increasing concentration of people. This situation should be the reversed for street prostitution, which, we conjecture, should show superlinear scaling. Buying sex from escorts, unlike being an escort, probably does not compete much with other activities, which explains the neutral scaling. Just as for the tendencies in sexual activities, the behavior of sellers seems more structured than that of buyers, suggesting that the seller's situations shape the dynamics more strongly than do the demands of the buyers.

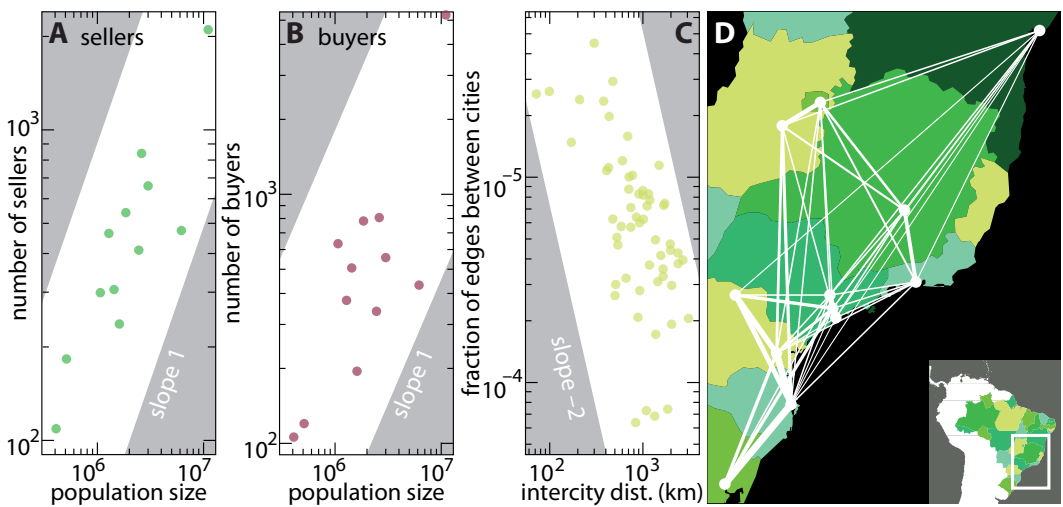

**Fig. 6.** (*A*) The scaling of the number of sellers in a city as a function of its population size. The shaded areas indicate isometric scaling (slope 1 in a double logarithmic plot). A fit to a power-law gives the exponent 0.72 ± 0.12 ($R^2$-value 0.78). (*B*) The corresponding plot for buyers with an exponent of 0.88 ± 0.21 ($R^2$-value 0.64). (*C*) Distance and the fraction of edges between two cities (normalized by the number of expected edges). The line indicates an inverse square-law. (*D*) The spatial network at the top of the map of Brazil, showing the connections between the sampled cities. The thickness is proportional to the number of edges between the two cities.

To investigate this geographic factor, we plotted the fraction of connections between individuals in different cities as a function of the distance between the cities (Fig. 6*C*). There is a negative correlation, reflecting travel patterns (cf. Ref. [29]) and not inconsistent with the inverse-square law found in trade patterns [30]. The vast majority of reported sex took place in the home cities of the buyers and sellers. The rare contacts between the cities become crucial bridges in the system-wide dynamics of both disease [29, 31] and information flow. The map of spatial network (Fig. 6*D*) illustrates that cities in southern Brazil are closer and more strongly connected (thicker lines) in comparison to the northern part. The central cities have larger populations in comparison to the others; whereas the southern cities have, in general, higher per capita incomes and human development indexes than the northernmost city. Statistics for the network at the end of sampling can be found in Table 1.

## DISCUSSION

Prostitution is usually a hidden, stigmatized socioeconomic phenomenon [32,33]. Its manifestations and social status has varied much throughout space and time. Nowadays, the escort business is increasingly coordinated over Internet in a way not too different from e.g. Internet dating services [34]. In this paper we have studied a moderated forum for discussing and rating encounters with escorts. This type of information-sharing helps reducing the risk for the sex-buyer, both of being disappointed with the sexual encounter and scammed. We have analyzed this dataset as a dynamic network and focused mostly on the information transfer at a moderated forum for discussing and rating encounters with escorts. We found a feedback from the postings (especially the grades given to the sex workers) to network evolution as well as a clustered, intermittent posting behavior, somewhat similar to emergent scaling in other patterns of human dynamics [6–8]. A high grade was a good predictor of high future degree. Whether the grade was medium or low did not seem to matter much. As an effect of the feedback from the forum to the patterns of sexual activity, we observed preferential attachment—that the rate of new contacts was proportional to the present degree to a positive power $\delta$. The preferential attachment for sex-sellers was close to linear ($\delta = 1$) at shorter time scales, but sublinear for longer sampling windows. This, we believe, reflects that sellers quit being escorts and are therefore no longer represented in the data. The resulting network had a broad, but narrower than power-law, degree distribution, negative degree correlations (disassortativity), and higher density of four-cycles than expected. Compared to scale-free networks (random networks only constrained by a power-law degree distribution) [13, 23], these three factors are all suggested to decelerate the spread of disease and information [12, 25, 27]. Also, related to disease spreading, we observe that some more risky sexual behaviors, such as anal or sex without condom, increase with the time of that the sellers or buyers are present in the data. This could perhaps be related to other studies that find correlations between the use of the Internet to find sex partners and risky sexual behavior [35]. The high density of short cycles is related to the geographic clustering of the network—cities define dense sub-networks; still, most of the network is connected into a giant component. The tendency for an intercity encounter decays in a way that is compatible with the inverse square-law observed in trade patterns. At the very least it is closer to an inverse-square law [30] than the exponential decrease observed in scientific collaborations [36]. The number of sellers scales sublinearly with the population of the city they are active in. This is in contrast to all other observed human endeavors that require people to meet [28] and adds an item to the list of odd features of prostitution. Our explanation is that the Internet-mediated escort business benefits less from an aggregation of people than do other competing activities, so it primarily takes market shares from e.g. street prostitution in smaller cities.

In summary, Internet-mediated prostitution (at least the Brazilian, high-end escorts in our data) is a phenomenon shaped both online and offline. In that sense it is similar to other social Internet services, like dating communities [5, 8] or even e-mail communication [7, 9]. A difference is that, in our data, the actual buyer-to-seller contact happens through external channels (primarily telephone), while in studies of other communication systems such direct contacts are usually recorded in the data. Many of the large-scale structures (broad degree-distributions, broad interevent-time distributions, long-time temporal correlations) can presumably be attributed to the same behavioral mechanisms as in those other communication systems (e.g. prefer-ential attachment [13, 23] as an explanation for heavy-tailed degree distributions, or queuing mechanisms [6] and the interplay between periodic behavior and cascade effects [37] as mechanisms behind heavy-tailed interevent-time distributions). However, our observed macro structures do not entirely match the observations from these other types of communication data (the preferential attachment is sublinear, there is a characteristic time scale of 100 days for the onset of long-term correlations, etc.). The peculiar economic [33, 34] and social [14] aspects of prostitution can presumably explain these deviations, but exactly how is a question for future research. It is hard to generalize our observations to prostitution in general. Some aspects, like the increasing frequency of more risky sexual activities (among both sellers and buyers), could perhaps be observed offline, while the feedback effects from communication between buyers are more pronounced in Internet-mediated prostitution than it could be offline.

## Acknowledgments

The authors are grateful to Luana de Freitas Nascimento, Devon Brewer and Diego Rybski for comments. FL acknowledges Riksbankens Jubileumsfond for financial support. PH acknowledges financial support from the Swedish Foundation for Strategic Research, Swedish Research Council and the WCU program through NRF Korea funded by MEST (R31–2008–000–10029–0).

## SUPPLEMENTARY INFORMATION

In this Supplementary Information we present further details relating to the network dynamics of the community.

### Cycles in the posting behavior

The community we study is, of course, not decoupled from the rhythms of life. In Fig. SI1, we plot a power-spectrum of the posts (or rather the occasions of the posts). Peaks in this plot correspond to cycles of a certain frequency. We see one major long-term peak at ~0.14 days$^{-1}$, corresponding to one week. There are probably 24 hours (circadian) rhythms too, but invisible due to our one-day resolution. The weekly, rhythm was not seen in a study of an Internet dating community [SI1], but it was observed in e-mail exchange [SI1] and Internet traffic [SI2].

### Evolution of number of posts

In Fig. SI2, we see that the number of posts by both the sex-buyers and sex-sellers are superlinear (indeed close to exponential) functions of the time elapsed between the first and last post $\Delta T$. This means that the sex-buyers who have been active the longest are also those who post most often. Like other reported phenomena, this effect may result from positive feedback, where some members become more active with time. The most conspicuous outliers among the sex-workers have a strength-value (number of posts about them) of about 80 posts over around 200 days. These outliers suggest a scenario in which some sex-workers are much requested at some point, but then stop working for some period of time. Compared to the sex-workers, buyers deviate less from the trend, with few members being very active over only a short period of time.

### Degree distribution

Assuming an attachment kernel of the form $k^\delta$ where $\delta$ is a positive constant, we find a sublinear preferential attachment. According to Ref. SI3, this would lead to a stretched exponential degree distribution. In Fig. SI3, we fit the degree distribution to a stretched-exponential form and, except for the lowest degrees, we find a good fit, and a fairly good validation of our assumption.

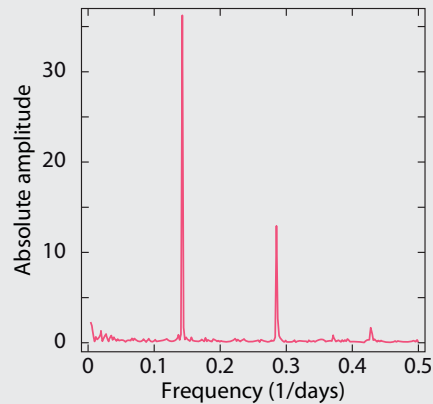

**Fig. SI1.** The power spectrum of the posts. The largest peak corresponds to one-week cycles (the following two peaks are corresponds to multiples of one week, i.e. they do not carry more information than the first peak).

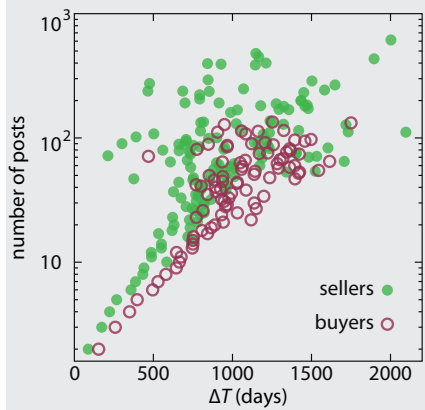

**Fig. SI2.** The total number of posts (about a sex-seller or that a sex-buyer has posted) as a function of the time in the community (the data is logarithmically binned along the abscissa).

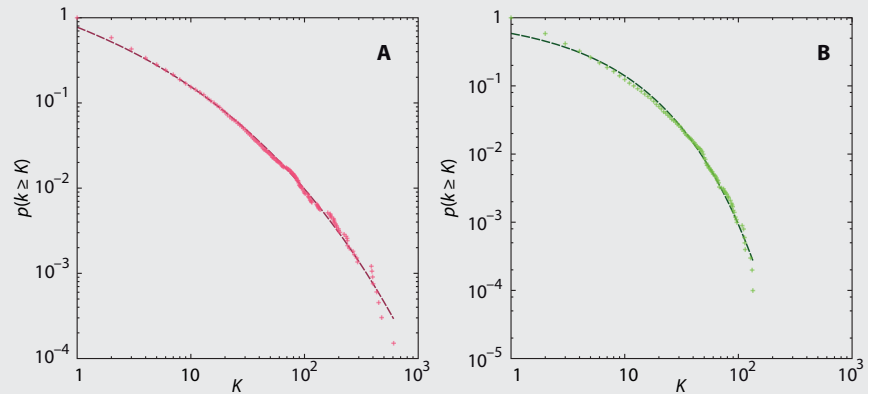

**Fig. SI3.** Cumulative degree distributions (same as in Fig. 3 in the paper) and stretched exponentials $A \exp(-(B K)^C)$ fitted with the Levenberg–Marquardt method. (A) shows the curve for sellers, (B) the curve for buyers. We obtain the parameter values (sellers) $A = 7.2 \pm 1.5$, $B = 30 \pm 12$, $C = 0.24 \pm 0.01$; and (buyers) $A = 1.0 \pm 0.2$, $B = 0.36 \pm 0.06$, $C = 0.55 \pm 0.02$.

**Strength distribution**

In Fig. 3 in the main text, we examined the degree distribution of the two types of nodes. As the degree does not count repeated contacts, it does not fully capture the activity of the sex-buyers and sex-sellers. In Fig. SI4 we plotted the distributions of strength, the total number of encounters reported by a sex-buyer, or about a sex-seller. We know from Table 1 in the paper that most contacts happen only once. It is thus not very surprising that the strength distribution is similar to the degree distribution. Just like the inter-event-time distribution [SI4, SI5] and degree distribution, the functional form can be described by a power-law with exponential cutoff. The cutoff is lower for buyers than sellers, which probably has many contributing reasons, one being economic: a larger strength value means higher income for the sex-sellers but greater cost for the sex-buyers.

**SI References**

1. Holme P (2003) Network dynamics of ongoing social relationships. *Europhys. Lett* 64:427–433.
2. Liu B, Fox EA (1998) Web traffic latency: Characteristics and implications. *Journal of Universal Computer Science* 4:763–778.
3. Krapivsky PL, Redner S, Leyvraz F (2000) Connectivity of Growing Random Networks. *Phys Rev Lett* 85:4629–4632.
4. Barabási A-L (2005) The origins of bursts and heavy tails in human dynamics. *Nature* 435:207–211.
5. Hidalgo CA (2006) Scaling in the inter-event time of random and seasonal systems. *Physica A* 369:877–883.

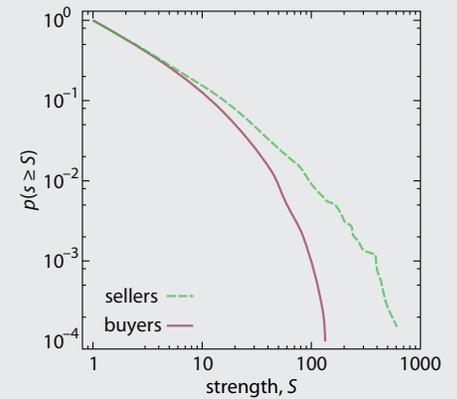

**Fig. SI4.** Cumulative distribution of the number of posts that a sex-buyer has posted, and in which a sex-seller is posted about.